\documentclass[
	pra,
	aps,
	reprint,
	amsmath,amssymb,
	a4paper,
	superscriptaddress,
	10pt
]{revtex4-2}
\usepackage{graphicx}
\usepackage{physics}
\graphicspath{{fig/}}
\usepackage[dvipsnames,table,xcdraw]{xcolor}
\definecolor{rmpblue}{HTML}{2e3092}
\usepackage[
	colorlinks=true,
	citecolor=rmpblue,
	linkcolor=rmpblue,
	filecolor=magenta,
	urlcolor=rmpblue,
]{hyperref}
\usepackage{upgreek}

\usepackage{siunitx}
\usepackage[version=3]{mhchem} 
\usepackage{textcomp}

\newcommand{\um}{\ensuremath{\,\upmu\text{m}}}
\usepackage{amsfonts}
\graphicspath{{fig/}}

\usepackage{physics}

\usepackage[dvipsnames,table,xcdraw]{xcolor}

\hypersetup{
    colorlinks=true,
    citecolor=blue,
    linkcolor=blue,
    filecolor=magenta,      
    urlcolor=blue
}

\newcommand{\iu}{\mathrm{i}\mkern1mu}
\newcommand{\eu}{\mathrm{e}\mkern1mu}

\newcommand{\ort}[1]{\boldsymbol{\mathbf{\hat{#1}}}}

\newcommand{\affilANU}{Nonlinear Physics Center, Research School of Physics, Australian National University, Canberra ACT 2601, Australia}

\newcommand{\affilIndia}{Department of Physics, Indian Institute of Technology Bombay, Mumbai 400076, India}

\newcommand{\affilIstanbul}{Department of Physics, Faculty of Science, Istanbul University, Vezneciler, 34134, Istanbul, Turkey}

\newcommand{\affilUW}{Department of Biomedical Engineering, University of Wisconsin-Madison, Madison, Wisconsin 53706, USA}

\begin{document}

\title{Intrinsic chirality of dielectric metasurfaces unlocked by resonant chiral modes}

\author{Brijesh Kumar}
\altaffiliation{Contributed equally}
\affiliation{\affilANU}
\affiliation{\affilIndia}

\author{Pavel Tonkaev}
\email[]{pavel.tonkaev@anu.edu.au}
\altaffiliation{Contributed equally}
\affiliation{\affilANU}

\author{Ivan Toftul}
\affiliation{\affilANU}

\author{Yihong Chen}
\affiliation{\affilUW}

\author{Vaishakh Unnikrishnan}
\affiliation{\affilUW}

\author{Albert Mathew}
\affiliation{\affilANU}

\author{Anshuman Kumar}
\affiliation{\affilIndia}

\author{Furkan Kuruoglu}
\affiliation{\affilUW}
\affiliation{\affilIstanbul}

\author{Filiz Yesilkoy}
\affiliation{\affilUW}

\author{Yuri Kivshar}
\email[]{yuri.kivshar@anu.edu.au}
\affiliation{\affilANU}

\begin{abstract}
Controlling optical chirality at the subwavelength scales is essential for many applications of nanophotonic structures in polarization optics, sensing, and nonlinear photonics. Achieving a strong chiroptical response in planar dielectric metasurfaces without intrinsically chiral building blocks (or "meta-atoms") remains challenging. The recent theoretical study [ACS Photonics {\bf 12}, 6717 (2025)] predicted that bilayer metasurfaces with rotated C$_4$-symmetric apertures can exhibit pronounced chiral response originating from resonant chiral photonic modes realizing maximum chirality under the mode strong coupling. That observation uncovers {\it a novel mechanism of metasurface chirality}. Here, we confirm experimentally this novel concept and demonstrate resonantly enhanced circular dichroism in the near-infrared frequency range. We fabricate a free-standing silicon membrane metasurface that is nominally achiral. When out-of-plane symmetry is broken by a thin PMMA layer, it unlocks and activates a strong chiral response. The observed circular dichroism is explained by the properties of chiral photonic modes, and it is governed by interlayer coupling and symmetry breaking, in agreement with theoretical predictions. These results establish bilayer metasurfaces as a simple and versatile platform for engineering strong mode-induced chirality in compact planar photonic metadevices.
\end{abstract}

\maketitle

Chirality is a fundamental concept in physics describing objects whose structures are not super-imposable on their mirror images and spanning systems from molecular to macroscopic scales~\cite{Barron2012Chirality}. Such systems exist in left- and right-handed enantiomer forms, which can exhibit distinct physical and chemical properties~\cite{prelog1976chirality, vukusic2009evolutionary, Deng2024NanoPhot}. In optics, chirality manifests itself as a differential response to left- and right-circularly polarized (LCP and RCP) light, giving rise to a phenomenon collectively referred to as optical chirality. The principal quantities used to characterize this effect are optical rotatory dispersion~\cite{eyring1968optical,castiglioni2011experimental} and circular dichroism (CD), defined as the difference in absorption or transmission between opposite circular polarizations~\cite{rodger1997circularDichroism,berova2000circularDichroism}.

In natural materials, chiral responses are typically weak, resulting in very small CD values. To overcome this limitation, artificially engineered structures such as metasurfaces have been extensively explored, enabling strong electric field localization and precise control over light-matter interaction~\cite{qiu2021quo,tonkaev2022all,kuznetsov2024roadmap}, as well as significantly enhanced optical chirality. Early developments largely relied on plasmonic structures~\cite{Yu2016plasmonicMSchiral,Ouyang:2018:optexp:plasmonic,Wang2019plasmonicMSchiral,Khaliq2023:AOM:Plasmonic:and:dielectric}, where strong field confinement is accompanied by inherent dissipative losses. More recently, all-dielectric nanostructures have emerged as a powerful alternative, capable of supporting low-loss resonances and achieving even stronger chiroptical responses through tailoring photonic eigenmodes~\cite{Solomon2019:ACSPhot:dielectricchiral,Tanaka:acsnano:2020:dielctric,Beutel2021:alldielectric,Khaliq2023:AOM:Plasmonic:and:dielectric,Wu:23:dielectric:chiral}. In particular, the concept of chiral quasi-bound states in the continuum has provided a powerful framework for selectively exciting resonant modes of a given handedness, enabling the realization of near-maximum optical chirality, where one circular polarization is efficiently transmitted while the other is strongly suppressed~\cite{Gorkunov2020PRL, dixon2021self,Gorkunov2021:BIC,overvig2021chiral,chen2023:BICchiral}.

A key practical consideration in planar photonic systems is the role of substrates, which often break out-of-plane mirror symmetry and thereby influence the overall chiral response. Recent theoretical studies have shown that even geometrically achiral metastructures can exhibit strong intrinsic optical chirality when placed on low-index substrates~\cite{Gorkunov2025AOM} or when a second layer is introduced in the free-standing membrane systems~\cite{Kumar2025chiral}, owing to engineered coupling between modes of different spatial parity. Ultrathin membrane platforms have only recently become available and have already enabled a wide range of applications, including harmonic generation~\cite{hallman2025high, tonkaev2025unconventional,tonkaev2025nonlinear}, optical sensing~\cite{adi2024trapping, rosas2025enhanced}, compact light-modulation devices~\cite{brikh2025mid} and enhanced emission from 2D materials~\cite{deng2026atomic}. These advances highlight the potential of membrane-based metasurfaces as a flexible platform for tailoring light-matter interactions at the nanoscale.

\begin{figure*}
    \centering
    \includegraphics[width=0.95\linewidth]{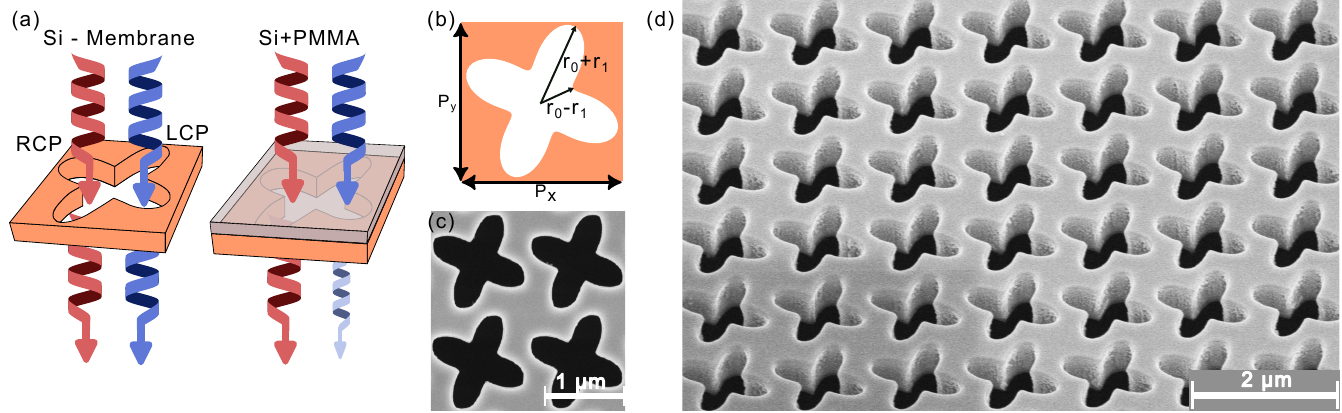}
    \caption{ \textbf{Chiral bilayer membrane metasurface}. (a) Schematic illustration of the designed metasurface structure. (b) Geometry of the unit cell used in the metasurface design. We used periodicity $P_x =P_y = P =1.2\um$, with $r_0 = 0.35P$ and $r_1 = 0.15P$ and parametric equation \eqref{eq:parametric} with $\theta = \pi/8$. (c) Top-view scanning electron microscopy (SEM) image showing four periodic unit cells of the fabricated metasurface. (d) Tilted SEM image recorded at a $45^{\circ}$ viewing angle, providing a larger-area view of the fabricated structure.}
    \label{fig1:Schamtic and fab}
\end{figure*}

In this work, we extend these concepts to bilayer dielectric free-standing membrane metasurfaces and provide their experimental realization and characterization. The studied free-standing membrane metasurfaces consist of square lattices of $\mathrm{C}_{4}$-symmetric apertures supporting pronounced resonant modes in the near-infrared range. While a single-layer silicon membrane exhibits an achiral optical response, the addition of a thin PMMA layer breaks the out-of-plane symmetry, enabling CD that varies from $-0.4$ to $0.4$ across different resonant modes in experiment. The agreement between theory and experiment demonstrates that the interplay of symmetry breaking and mode hybridization provides a powerful mechanism for achieving CD. Our results establish a general route to achieving strong chiroptical responses in planar, nominally achiral systems, opening new opportunities for polarization control, chiral sensing, and nonlinear photonics applications.

Our structure consists of a square lattice of rotated $\mathrm{C}_4$-symmetric apertures patterned into a free-standing silicon membrane. Figure~\ref{fig1:Schamtic and fab}(a) shows a schematic of the metasurface configuration, where the second layer breaks out-of-plane mirror symmetry, enabling chiroptical effects. Figure~\ref{fig1:Schamtic and fab}(b) depicts the unit-cell geometry. The structure is defined by the parametric equation
\begin{equation}
    r = r_0 + r_1 \cos[4(\phi + \theta)],
    \label{eq:parametric}
\end{equation}
where $r_0 = 0.35P$, $r_1 = 0.15P$, and $\theta = \pi/8$. The lattice is periodic in both the $x$- and $y$-directions with periodicity $P_x = P_y = P = 1.2\um$. The free-standing membrane metasurfaces were fabricated using electron beam lithography (see Section S1 in the Supporting Information). Figure~\ref{fig1:Schamtic and fab} (c) and (d) present scanning electron microscopy (SEM) images of the fabricated single-layer metasurfaces, including a top-view of four unit cells and a large area view confirming the high fabrication quality and uniformity across extended regions.

To achieve strong CD, our design exploits symmetry conditions and corresponding selection rules. To characterize our system, we use the following definition of the linear CD:
\begin{equation}
    \mathrm{CD}_{\mathrm{co}} = \frac{T_{\mathrm{LL}} - T_{\mathrm{RR}}}{T_{\mathrm{LL}} + T_{\mathrm{RR}}},
    \label{eq:CDco}
\end{equation}
where $T_{\mathrm{RR}}$ and $T_{\mathrm{LL}}$ denote the transmitted intensities for RCP and LCP light, respectively, under corresponding circularly polarized excitation. The presence of any mirror symmetry plane ensures $\mathrm{CD}_{\mathrm{co}} = 0$~\cite{Koshelev2024JOPT}. To achieve the maximum optical chirality of the metasurface, the metasurface modes must be carefully engineered~\cite{Kumar2025chiral}. 

In particular, optical chirality can be achieved via hybridization of opposite parity modes, and can be predicted through a mode-resolved CD defined as~\cite{Gorkunov2025AOM,Toftul2024PRL,Kumar2025chiral}
\begin{equation}
	\mathrm{CD}_{\text{mode}, n} = \frac{|m_{n\mathrm{L}} m^{\prime}_{n \mathrm{L}}|^2 - |m_{n\mathrm{R}} m^{\prime}_{n\mathrm{R}}|^2}{|m_{n\mathrm{R}} m^{\prime}_{n \mathrm{R}}|^2 + |m_{n\mathrm{L}} m^{\prime}_{n\mathrm{L}}|^2}.
\end{equation}
Here, $m_{n\mathrm{R}}$, $m_{n\mathrm{L}}$ describe the coupling of the $n$-th eigenmode to RCP and LCP waves propagating in the $-z$ direction, while $m'_{n\mathrm{R}}$ and $m'_{n\mathrm{L}}$ correspond to the opposite $+z$ direction. These coupling coefficients are given by
\begin{align}
	\begin{split}
		m_{n \mathrm{R}} &= A_n \int \limits_{\mathbb{V}_{\text{M}}}  \left[\varepsilon(
		\vb{r})  - 1\right]\mathbf{E}_n(\mathbf{r}) \cdot \ort{e}_{+} \eu^{-\iu \Omega_n z/c} \, \dd V, \\
		m_{n \mathrm{L}} &= A_n \int \limits_{\mathbb{V}_{\text{M}}}  \left[\varepsilon(
		\vb{r})  - 1\right]\mathbf{E}_n(\mathbf{r}) \cdot \ort{e}_{-} \eu^{-\iu \Omega_n z/c} \, \dd V,
	\end{split}
	\label{eq:m}
\end{align}
while those for waves propagating in the opposite $+z$ direction read as:
\begin{align}
	\begin{split}
		m^{\prime}_{n \mathrm{R}} &= A_n \int \limits_{\mathbb{V}_{\text{M}}}  \left[\varepsilon(
		\vb{r})  - 1\right]\mathbf{E}_n(\mathbf{r}) \cdot \ort{e}_{-} \eu^{\iu \Omega_n z/c} \, \dd V, \\
		m^{\prime}_{n \mathrm{L}} &= A_n \int \limits_{\mathbb{V}_{\text{M}}}  \left[\varepsilon(
		\vb{r})  - 1\right]\mathbf{E}_n(\mathbf{r}) \cdot \ort{e}_{+} \eu^{\iu \Omega_n z/c} \, \dd V,
	\end{split}
	\label{eq:m_prime}
\end{align}
where circular polarization basis vectors $\ort{e}_{\pm} = (\ort{x} \pm \iu \ort{y})/\sqrt{2}$, weighted by the permittivity contrast $\varepsilon(\mathbf{r}) - 1$ over the metasurface volume $\mathbb{V}_{\text{M}}$, and include plane wave phase factors $\exp(\mp \iu \Omega_n z/c)$, where $\Omega_n = \omega_n - \iu \gamma_n$ are complex eigenfrequencies. The factors $A_n$ depend on the mode normalization and take a simple form $A_n=\iu \Omega_n \sqrt{\varepsilon_0/(2\mathbb{A}c)}$, where $\mathbb{A}$ is the area of metasurface unit cell and $c$ is the speed of light in vacuum~\cite{Gorkunov2025AOM}.

\begin{figure*}
    \centering
    \includegraphics[width =\linewidth]{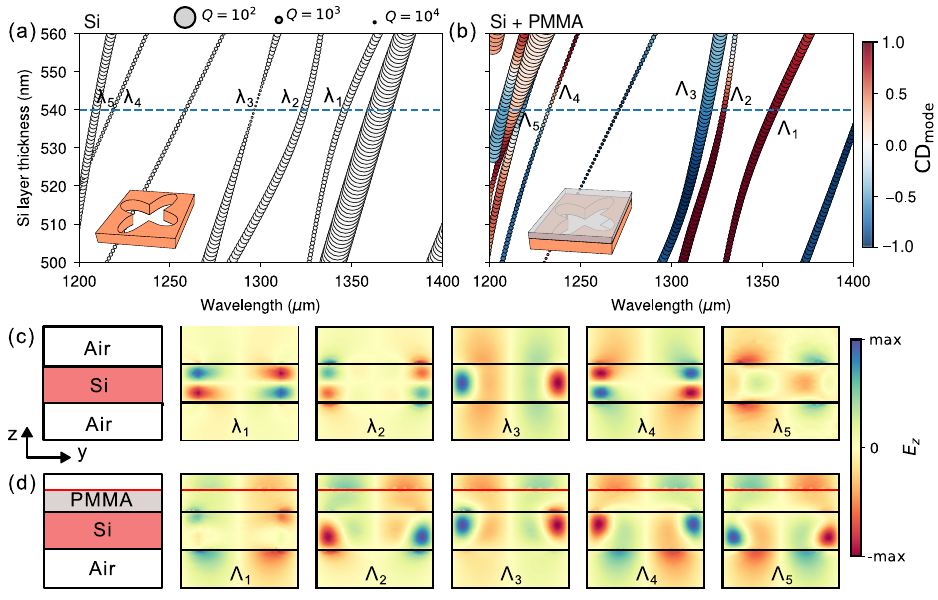}
    \caption{\textbf{Mode coupling via symmetry breaking.} 
		Variation of eigenfrequencies as a function of silicon layer thickness in membrane metasurfaces with a unit cell periodicity of 1.2\um. (a) Eigenfrequencies as a function of silicon layer thickness for a 2D chiral metasurface. (b) Eigenfrequencies as a function of bottom silicon layer thickness for 3D chiral membrane metasurfaces, where the top layer (refractive index $n = 1.47$) has a thickness of 300\,nm. The $z$-component of the electric field ($E_z$) of the eigenmodes in the $yz$-plane at the center of the unit cell ($x$=0) is shown for (c) a single-layer membrane metasurface and (d) a double-layer membrane metasurface with a silicon thickness of 540\,nm.} 
    \label{fig2:mode hybrid}
\end{figure*}

To engineer the metasurface eigenmodes for achieving maximum optical chirality, we employ a design strategy guided by our previous theoretical work~\cite{Kumar2025chiral}. In the experimentally fabricated membranes, the silicon thickness is typically less precisely controlled than in-plane geometrical parameters. To account for this practical tolerance, we simulated the eigenmode evolution for a single-layer silicon membrane with thicknesses ranging from 510\,nm to 560\,nm, focusing around the fabricated value of $540\pm5$\,nm. We select high-fidelity fabricated structures with the lattice period $P = 1.2\um$ to identify the regime of strong mode hybridization.

\begin{figure*}
   \centering
    \includegraphics[width=0.9\linewidth]{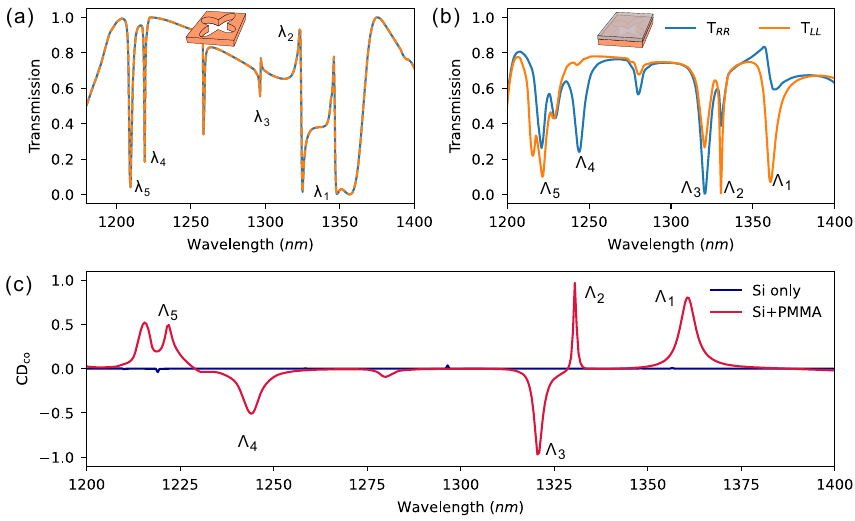} 
    \caption{\textbf{Numerically calculated transmission.} (a) Co-polarized transmission for RCP and LCP light in a single-layer silicon membrane metasurface. (b) Co-polarized transmission for RCP and LCP light in a bilayer Si+PMMA membrane metasurface. (c) Circular dichroism of the single- and bilayer membrane metasurfaces.}
    \label{Fig3:Numerical transmission}
\end{figure*}

Figure~\ref{fig2:mode hybrid}(a) shows the calculated evolution of eigenmodes as a function of silicon thickness for this fixed periodicity. The size of each circle represents the imaginary part of the eigenfrequency, corresponding to inverse quality factor ($Q$-factor). The results reveal several mode hybridization regions. For the single-layer silicon membrane, the modes are labeled as $\lambda_{1-5}$. Figure~\ref{fig2:mode hybrid}(b) presents the same analysis after introducing a 300\,nm top layer of PMMA with refractive index $n= 1.47$. The corresponding hybridized modes are denoted as $\Lambda_{1-5}$. 
In the single-layer metasurface, modes $\lambda_1$, $\lambda_2$ are already coupled due to their identical parity, as confirmed by the $E_z$-component of the $E$-field shown in Figure~\ref{fig2:mode hybrid}(c). In the vicinity of this anti-crossing there is a third mode $\lambda_3$ of opposite parity. This gives a foundation of a peculiar three-mode interaction scenario described in details in Ref.~\cite{Kumar2025chiral}. But no chirality emerges because the structure retains out-of-plane mirror symmetry. Mode crossing between same parity modes $\lambda_4$, $\lambda_5$ occurs near a thickness of 530\,nm. These modes do not hybridize because of their opposite parity as per Figure~\ref{fig2:mode hybrid}(c). Introducing the second dielectric layer breaks the out-of-plane symmetry and enables parity mixing, leading to  hybridization among $\Lambda_1$, $\Lambda_2$, $\Lambda_3$. Notably, modes $\Lambda_1$, $\Lambda_2$ preferentially couple to one circular polarization, whereas $\Lambda_3$ couples to the opposite one. A similar hybridization behavior is also observed for modes $\Lambda_4$, $\Lambda_5$, however some extra modes appear near the crossing with the addition of the PMMA layer. These results establish a direct correlation between the mode parity and the handedness of the chiral response. In the simulations, the PMMA thickness was set to 300\,nm, consistent with the experimental fabrication conditions, where a 2\% PMMA solution in anisole yields a minimum thickness of approximately 280\,nm after spin-coating during the electron-beam lithography resist preparation process.

In the fabricated membrane metasurfaces, the lattice periodicity is $1.2\um$, and the membrane thickness is $540 \pm 10$\,nm. Using these experimentally relevant parameters, we calculate the transmission spectra, as shown in Figure~\ref{Fig3:Numerical transmission}. Figure~\ref{Fig3:Numerical transmission}(a) presents the transmission spectra of the single-layer (silicon only) membrane metasurface. The resonances labeled $\lambda_{1-5}$ correspond to the same modes identified in Figure~\ref{fig2:mode hybrid}(b), confirming consistency between modal analysis and spectral response.
The considered structures consist of square lattices of $\mathrm{C}_{4h}$-symmetric holes, which naturally support degenerate resonant modes with well-defined spatial parity. In the single-layer configuration, the presence of an out-of-plane mirror symmetry plane makes the system geometrically achiral, thereby forbidding CD despite the presence of pronounced optical resonances. By introducing a second dielectric layer, the symmetry is reduced from $\mathrm{C}_{4h}$ to $\mathrm{C}_4$, which breaks the mirror symmetry while preserving rotational invariance and making the system geometrically chiral~\cite{Sinev2025}. This symmetry reduction enables parity mixing and induces chiral hybridization of the photonic modes.

Building on the mode analysis above, Figure~\ref{Fig3:Numerical transmission}(b) shows the numerically calculated transmission of the bilayer membrane metasurface, where a top layer with refractive index $1.47 + 0.02\iu$ is introduced. The imaginary part is treated as an effective phenomenological loss that lumps together fabrication-related dissipation channels, such as polymer granularity, sidewall roughness, structural disorder, and out-of-plane scattering, rather than the extinction coefficient of bulk PMMA, which is $\sim 10^{-5}$ in the near-infrared~\cite{Beadie2015ApplOpt,zhang2020complex}. The value $0.02$ was chosen as a representative magnitude that yields CD of the same order as the measurement. Per our previous theory~\cite{Kumar2025chiral}, the sign and spectral structure of CD are set by mode symmetry, while only its amplitude depends on the loss level. The resonances labeled $\Lambda_{1-5}$ correspond to the hybridized modes identified in Figure~\ref{fig2:mode hybrid}(b), demonstrating how the bilayer architecture reshapes both the modal landscape and the spectral positions of the resonances. Near the crossing associated with $\lambda_{4-5}$, additional resonances emerge, resulting in pronounced mode crowding and further hybridization. Consequently, strict sequential mode numbering becomes less well-defined in this region.

\begin{figure*}
    \centering
    \includegraphics[width=0.9\linewidth]{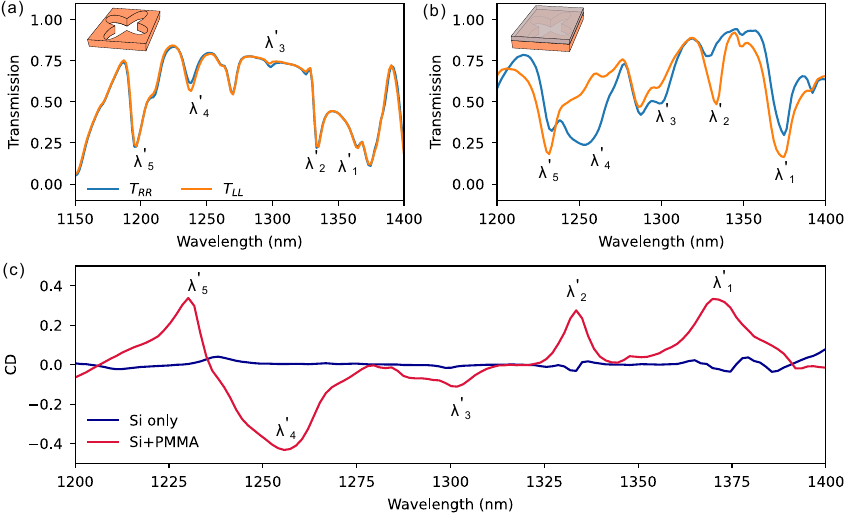}
    \caption{\textbf{Experimentally measured transmission.} (a) Co-polarized transmission for RCP and LCP light measured for a single-layer silicon membrane metasurface. (b) Co-polarized transmission for RCP and LCP light measured for a silicon membrane covered by a PMMA layer. (c) Experimental circular dichroism of the single- and bilayer membrane metasurfaces.}
    \label{fig4:Exp}
\end{figure*}

Figure~\ref{Fig3:Numerical transmission}(c) presents the calculated CD in transmission ($\mathrm{CD}_{\mathrm{co}}$), as defined in Eq.~\eqref{eq:CDco}. Owing to reciprocity, rotationally symmetric structures under normal incidence cannot exhibit transmission CD in the absence of dissipation, even when geometric chirality is present. To overcome this limitation, we introduce controlled losses in the spectral regions where chiral mode hybridization is strongest. Under these conditions, interference between nearly degenerate chiral resonances becomes highly sensitive to dissipative channels, producing different attenuation for LCP and RCP excitation. As a result, a pronounced transmission CD emerges. Notably, both the sign and magnitude of CD are governed by the symmetry and parity of the interacting modes. A direct comparison between the mode-resolved circular dichroism ($\mathrm{CD}_{\text{mode}, n}$) and the transmission-derived circular dichroism ($\mathrm{CD}_{\mathrm{co}}$) reveals that the sign of $\mathrm{CD}_{\mathrm{co}}$ at each resonance matches the dominant circular polarization of the corresponding mode for all three identified resonant modes ($\Lambda_{1-3}$). This agreement confirms that the chiral spectral response originates directly from symmetry-enabled mode hybridization and the parity structure of the underlying photonic eigenmodes.

To validate the numerical predictions, we next experimentally characterize the fabricated metasurfaces under the same observables used in the simulations. The membrane metasurfaces were illuminated with low-NA light under RCP and LCP excitation, while the transmitted signal was analyzed into RCP and LCP components (see mode details in Section S2 of Supporting Information). Figure~\ref{fig4:Exp} presents the measured transmission spectra together with the corresponding CD. This one-to-one comparison with numerical results allows us to assess the fidelity of the model and to verify the physical origin of the observed chiroptical response.

Figure~\ref{fig4:Exp}(a) shows the experimentally measured transmission spectrum of the single-layer (silicon only) membrane metasurface. The experimental spectral features closely reproduce the calculated response, with resonances appearing near the wavelengths associated with $\lambda_{1-5}$, consistent with the modes identified in Figure~\ref{fig2:mode hybrid}(a). Minor deviations in resonance positions and linewidths are attributed to fabrication tolerances, material inhomogeneity, and the finite numerical aperture of the optical system.

Figure~\ref{fig4:Exp}(b) presents the transmission spectrum of the bilayer membrane metasurface. Consistent with the numerical predictions, the introduction of the additional layer modifies the spectral response, leading to resonance shifts, linewidth changes, and altered mode contrast. The experimentally observed resonances $\lambda^\prime_{1-5}$ correspond to the hybridized modes discussed in Figure~\ref{Fig3:Numerical transmission}(b), directly confirming that interlayer coupling governs the reconstructed modal landscape and chiral optical response.

Finally, Figure~\ref{fig4:Exp}(c) shows the experimentally extracted circular dichroism, defined in Eq.~\ref{eq:CDco}. Pronounced CD features emerge at the resonance wavelengths, with CD signs that closely follow the numerical predictions. This agreement confirms that the measured chiroptical response originates from the same symmetry-breaking and modal interference mechanisms identified in the theoretical analysis. Overall, the experiments show strong qualitative agreement with the simulations, demonstrating the robustness of the bilayer metasurface platform for engineering resonant chiral light–matter interactions. Consistent with this interpretation, control measurements performed on the unpatterned membranes, including both the single-layer silicon and bilayer Si/PMMA structures, show no detectable chiral response due to their in-plane symmetry (see Section S3 in the Supporting Information).

A clear quantitative discrepancy is observed between the circular dichroism values obtained from the numerical simulations and those measured experimentally. This deviation is primarily attributed to the strong sensitivity of transmission chirality to optical dissipation, which plays a central role in enabling and enhancing circular dichroism~\cite{Kumar2025chiral}. In the fabricated metasurfaces, the effective loss channels can differ from the idealized parameters used in the simulations because of surface roughness, structural disorder, residual absorption, and out-of-plane scattering. Since the chiroptical response is governed by interference between near-degenerate hybridized modes, even modest variations in these dissipative pathways can lead to substantial changes in the magnitude of the measured circular dichroism. 

Our results establish bilayer dielectric metasurfaces as a versatile and scalable platform for engineering chiroptical responses, offering new opportunities for advanced polarization control, optical sensing, and integrated nanophotonics. Beyond the linear regime, such architectures are particularly promising for nonlinear chiroptical phenomena, including high-harmonic generation from free-standing membranes~\cite{tonkaev2025unconventional} and chiral third-harmonic generation in suspended dielectric platforms~\cite{tonkaev2025nonlinear}. The bilayer design provides an additional degree of freedom for tailoring modal overlap and symmetry, enabling enhanced efficiency and controllability in nonlinear photonic functionalities. Recent demonstrations of CMOS-compatible photonic platforms for excitonic valley state routing~\cite{KumarMandal2023}, together with monolithic integration of solid-state emitters and photonic resonators for efficient on-chip light-matter coupling~\cite{Mandal2024}, further suggest that combining such approaches with chiral metasurfaces could enable handedness-selective routing and compact multifunctional quantum photonic devices.

In conclusion, we have demonstrated that bilayer dielectric metasurfaces provide a powerful and scalable platform for engineering a strong chiroptical response through controlled hybridization of chiral modes. By combining numerical modeling and experimental measurements, we have established a direct correspondence between the underlying photonic eigenmodes and the observed transmission and circular dichroism spectra. The introduction of a secondary layer enables tunable intermode coupling, resulting in enhanced and controllable chirality even in systems with nominally symmetric geometries.  More broadly, our results highlight the central role of symmetries, losses, and interlayer interactions in shaping resonant chiral optical responses. The excellent agreement between numerical simulations and experimental data validates the theoretical framework and confirms the robustness of the proposed design strategy. The ability to tailor both linear and nonlinear chiroptical effects in scalable dielectric metasurfaces opens new opportunities for polarization-selective photonics, sensing, and integrated optical technologies. For example, the layer can be replaced by some van der Waals material, thus inducing strong coupling between the circularly polarized photons and spin-polarized (valley) excitons (bound electron-hole pairs). We believe our findings pave the way for advanced platforms capable of exploiting chiral light–matter interaction with high precision, compact footprint, and broad functional tunability.

\medskip
\textbf{Acknowledgements} \par 
Y.K. acknowledges a support from the Australian Research Council (Grant DP210101292) and the International Technology Center Indo-Pacific (ITC IPAC) via Army Research Office (contract FA520923C0023). Photonic metasurfaces were fabricated at the Center for Nanoscale Materials at Argonne National Laboratory, a U.S. Department of Energy Office of Science User Facility, supported by the U.S. DOE Office of Basic Energy Sciences under Contract No. DE-AC02-06CH11357. The authors thank Dr David A. Czaplewski from ANL for their support with the fabrication process. The authors thank Maxim Gorkunov for useful discussions.

\bibliography{ref}

\end{document}